\documentclass[preprint,showpacs,prl,amsfonts,amsmath,amssymb]{revtex4-1}
\usepackage{pifont}
\usepackage{graphicx}
\usepackage{amssymb}
\usepackage{dcolumn}
\usepackage{amsmath}
\usepackage{bm}

\usepackage{colordvi}
\usepackage{multirow}
\usepackage{array}
\usepackage{booktabs}
\usepackage[flushleft]{threeparttable}
\usepackage{sidecap}
\begin{document}

\title{Structural stability of lattice-matched heterovalent semiconductor superlattices}

\author{Hui-Xiong Deng}
\affiliation{State Key Laboratory for Superlattices and
Microstructures, Institute of Semiconductors, Chinese Academy of
Sciences, P. O. Box 912, Beijing 100083, China}
\affiliation{National Renewable Energy Laboratory, Golden, CO 80401, USA}
\author{Bing Huang}
\affiliation{National Renewable Energy Laboratory, Golden, CO 80401, USA}
\author{Jingbo Li}
\author{Shu-Shen Li}
\affiliation{State Key Laboratory for Superlattices and
Microstructures, Institute of Semiconductors, Chinese Academy of
Sciences, P. O. Box 912, Beijing 100083, China}
\author{Su-Huai Wei}
\email{swei@nrel.gov}
\affiliation{National Renewable Energy Laboratory, Golden, CO 80401, USA}


\begin{abstract}
Lattice-matched heterovalent alloys and superlattices have some unique physical properties. For example, their band gap can change by a large amount without significant change in their lattice constants, thus they have great potential for optelectronic applications. Using first-principles total energy calculation and Monte Carlo simulation as well as lattice harmonic expansion, we systematically study the stability of the heterovalent superlattices. We show that the chemical trend of stability of lattice-matched heterovalent superlattices is significantly different from lattice-mismatched isovalent superlattices, because for lattice-mismatched isovalent superlattices the stability is mostly determined by strain, whereas for lattice-matched nonisovalent superlattices the interfacial energy depend not only on the bond energy but also on the Coulomb energy derived from donor- and acceptor-like wrong bonds. In the short-period heterovalent superlattices, the abrupt [111] interface has the lowest energy even though it is polar, whereas for the long-period heterovalent superlattices, the [110] interface has the lowest energy. On the contrary, [201] superlattices are usually the most stable for lattice-mismatched isovalent superlattices.
\end{abstract}

\maketitle
\section{I. INTRODUCTION}
Mixing different elemental or binary semiconductor compounds to form alloys or superlattices is a common practice to extend available material properties for specific technological applications\cite{Alloy_book1,Jaros}. In the past, most of the studies of semiconductor alloys have been for common-anion AX and BX alloys A$_{1-x}$B$_{x}$X or for common-cation AX and AY alloys AX$_{1-x}$Y$_{x}$ because these common-atom alloys can be easily mixed and their material properties usually change smoothly with alloy composition $x$. Compared to common-atom alloys, the study of heterovalent semiconductor alloys (AX)$_{1-x}$(BY)$_{x}$ or superlattices (AX)$_{n}$(BY)$_{m}$ (e.g., AX and BY are the III-V and II-VI compounds, respectively), is still lacking, mainly because of their structural and chemical complexities. However, unlike common-atom alloys, heterovalent alloys and superlattices often have their own advantages. For example, for conventional common-atom alloys such as (Ga)$_{1-x}$(In)$_{x}$As, adding InAs to GaAs to reduce the band gap also simultaneously increases the lattice constant. The large strain between the alloys and their constituents makes it difficult to grow high-quality GaAs/GaInAs heterostructures, which are necessary for some devices such as high-efficiency tandem solar cells and light-emitting diodes. However, one can find many heterovalent systems\cite{Alloy_book2,RGDandrea1990} (Si, GaP), (GaAs, ZnSe), (GaSb, ZnTe), etc., that have lattice-matched constituents, so the band gap can change by a large amount as a function of alloy composition without significant change of the lattice constant. Due to this unique property of heterovalent systems, several studies have recently been carried out to explore their great potential for application in optoelectronic devices\cite{Lambrecht,Ohno,Nicolini,Maeda_JACS,colli_controlling_2003,Frey,Shuzhi_Wang}. For example, lattice-matched GaN/ZnO alloys have been suggested as a good candidate for electrodes for high-efficiency solar hydrogen production through photoelectrochemical water splitting\cite{Maeda_JACS,maeda_photocatalyst_2006,LWWang_PRL}, and lattice-matched ZnSe/GaAs and ZnTe/GaSb heterostructures are proposed to be high quality blue-green emitters\cite{haase_bluegreen_1991,Schulz}. ZnTe/GaSb also has great potential for tandem solar cell absorbers because GaSb has a direct band gap of 0.8 eV and ZnTe has a direct band gap of 2.4 eV; therefore, their alloys and superlattices are capable of covering a large range of the solar spectrum with negligible change in the lattice constant.

However, there are still many fundamental issues that need to be understood for these heterovalent systems: (i) In conventional common-atom isovalent alloys (e.g., GaAs/GaP\cite{RGDandrea} or GaAs/AlAs\cite{SHWei1988}) there exist only two types of bonds that are the same as their binary compounds with the concentration of the bond type uniquely determined by the alloy concentration $x$. However, for heterovalent AX/BY systems, there will inevitably be two additional types of bonds (i.e., A-Y and B-X, hereinafter referred to as wrong bonds) besides the ones that are the same as the binary compounds (A-X and B-Y, hereinafter referred to as right bonds), and the number of the wrong bonds and right bonds is not uniquely determined by the alloy concentration $x$. In principle, the wrong bonds will increase the alloy formation energy\cite{Shuzhi_Wang,LWWang_PRL}. Consequently, it is expected that the fewer wrong bonds in the alloy, the more stable for the nonisovalent structures. This suggests that one way to lower the formation energy of heterovalent systems is to form long-range superlattices instead of short-range disordered alloys; (ii) For heterovalent systems, the wrong bond is not saturated with two electrons per bond. For example, for the III-V/II-VI systems, each II-V wrong bond will have a deficiency of 1/4 of an electron, namely, providing 1/4 holes, while each III-VI wrong bond will have 1/4 excess electrons in tetrahedral coordinated structures. Because of this charge imbalance, there will be charge transfer from the electron-excess bond to the electron-deficiency bonds and the subsequent Coulomb interaction between the charged centers can lower the energy and is most efficient if the oppositely charged centers are close to each other, forming disordered or short-range ordered superstructures. Therefore, it is not clear whether the stable interfacial configurations of the heterosuperlattices should be polar or non-polar. Also unclear is the atomic configuration of the lowest-energy interface structures of the heterosuperlattices as a function of the superlattice orientation and layer thickness.

\begin{figure}[h]
\centering
\includegraphics*[width=10.0cm,keepaspectratio]{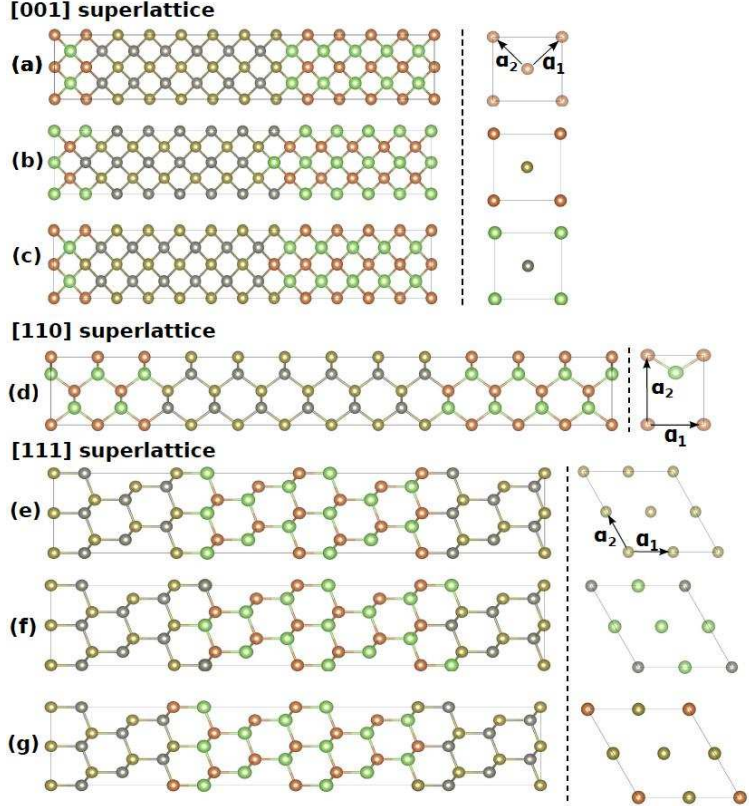}
\caption{\label{fig:f3}(Color online) Schematic structures of lattice-matched heterovalent (GaSb)$_{n}$/(ZnTe)$_{n}$ superlattices in forming (a) abrupt (AB), (b) mixed anion (MA), and (c) mixed cation (MC) interfaces along the [001] orientation, and (d) abrupt (AB) interfaces along the [110] orientation, and (e) abrupt (AB), (f) mixed cation (MC), and (g) mixed anion (MA) interfaces along the [111] orientation. To the right of each plot represents the atomic arrangements of the lateral dimensions near one interface. $a_{1}$ and $a_{2}$ are the basis vectors of the lateral plane for the corresponding abrupt interfaces. The green, orange, gray, and dark yellow balls indicate the Ga, Sb, Zn, and Te atoms, respectively.}
\end{figure}

In this paper, we select the lattice-matched heterovalent (GaSb)$_{n}$/(ZnTe)$_{n}$ superlattices as an example, and analyze the structural stability of the interfaces as a function of period \emph{n} along different growth orientations. Through Monte Carlo simulations combined with {\it ab initio} total-energy calculations, we find that: (i) For the short-period superlattices, the interfacial energy is mainly controlled by the interfacial bond energy, i.e., the number of wrong bonds per unit interfacial area. This explains why the [111] orientation has the polar abrupt (AB) interface when the period is small ($n<4$) and why [111] and [110] have lower energy than [001] and [201] interfaces; (ii) For the long-period superlattices,  the Coulomb interaction becomes more important in determining the stability of interfacial structures, so for long-period superlattices, nonpolar interfaces are more stable, which are achieved by intermixing of atoms in the interfaces. In this case, the [110] interface is the most stable orientation, because it is naturally nonpolar and has the fewest wrong bonds per unit interfacial area.

\subsection{II. METHOD OF CALCULATIONS}
The model structures studied in this paper are the lattice-matched (GaSb)$_{n}$/(ZnTe)$_{n}$ superlattices with $n$=1-6 along the [001], [110], and [111] directions, respectively. Results for other directions can be fitted by using lattice harmonic expansion. For the atomic arrangements parallel to the interface, we choose the ($\sqrt{2}\times\sqrt{2}$), (1$\times$1), and (2$\times$2) supercells, respectively, for [001], [110], and [111] superlattices. The Monte Carlo method is employed to search the stable interfacial structures by swapping the atoms near the interfaces. We only allow a cation swapping with a cation or an anion swapping with an anions in each trial Monte Carlo step. After each swapping, the total energies are calculated using density functional theory (DFT)\cite{DFT01,DFT02} within the generalized gradient approximations (GGA)\cite{GGA,PBE} as implemented in the Vienna {\itshape ab initio} simulation package (VASP) \cite{vasp01,vasp02,vasp03}. The electron and core interactions are included based on the frozen-core projector-augmented-wave approach\cite{PAW}. The cutoff energy for the wavefunction expansion is 450 eV. Convergence with respect to the Monkhorst-Pack $k$-point mesh in the Brillouin zone \cite{Monkhorst} has also been carefully checked to assure the error of interfacial energy is within 0.1 meV/\AA$^{2}$. In the relaxation process, all the atoms are allowed to relax until the quantum mechanical forces acting on them become less than 0.02 eV/{\AA}.

The interfacial energy is defined as:
\begin{equation}
\begin{aligned}
\gamma=\frac{E_{tot}[(AX)_{n}(BY)_{n}]-(nE_{tot}(AX)+nE_{tot}(BY))}{2S}\\
\end{aligned}
\end{equation}
where $E_{tot}[(AX)_{n}(BY)_{n}]$, $E_{tot}(AX)$, and $E_{tot}(BY)$ are the total energies of superlattices, bulk AX, and BY, respectively. $S$ represents the interfacial area.

\subsection{III. RESULTS AND DISCUSSIONS}
To compare the stability of different interface structures, we first try to find the superlattices with nonpolar interfaces, which are expected to be stable, at least for long-period superlattices. For the nonpolar interfaces, the interface should have equal donor and acceptor wrong bonds. For the [001] superlattices, our calculations find that the stable nonpolar mixed interfaces are constructed by mixing 50\% Ga and 50\% Zn atoms [mixed cation (MC)] or mixing 50\% Sb and 50\% Te atoms [mixed anion (MA)] at the interface [Figs. 1(b) and 1(c)], in agreement with previous predictions\cite{Kley,farrell_cation_2004,farrell_anion_2005}. This is because along the [001] orientations, the number of wrong bonds eliminated and created are always the same, so the total number of wrong bonds is not changed after the switched of two cations or two anions at the interfaces. For the [111] superlattices, each time a cation or an anion is switch to remove a wrong bond along the [111] direction, it will simultaneously create three wrong bonds away from the [111] direction with different types (acceptors or donors). Consequently, in order to form nonpolar interfaces only one atom should be switched out of every four [111] wrong bonds, which is consistent with our Monte Carlo simulations [Figs. 1(f) and 1(g)]. Finally, for [110] superlattices, the abrupt interface [Fig. 1(d)] already has an equal number of Ga-Te and Zn-Sb wrong bonds at each interface; therefore, it is already nonpolar.

\begin{table*}[!htbp]
\begin{threeparttable}
\caption{Detailed information for the abrupt (AB), mixed cation (MC), or mixed anion (MA) interfaces of (GaSb)$_{n}$(ZnTe)$_{n}$ superlattices along [001], [110] and [111] orientations. For the number of wrong bonds, the numbers outside and inside the parentheses are for the two interfaces of the superlattices, respectively.}
\centering
\begin{tabular}{cccccccc}
\toprule
Growth orientation &\multicolumn{3}{c}{[001]} &[110] &\multicolumn{3}{c}{[111]}\\
\hline
Interfacial characters& AB & MC & MA & ~~~~~~AB~~~~~~ & AB & MC & MA\\
\hline
Polar (P) or nonpolar (NP) & P    &NP       &NP      & NP     & P    &NP       &NP\\
\hline
Wrong bonds & &  & &&&&\\
Ga-Te & 4(0) & 2(2) & 2(2) & 1(1) & 4(0) & 3(3) & 3(3) \\
Zn-Sb & 0(4) & 2(2) & 2(2) & 1(1) & 0(4) & 3(3) & 3(3) \\
Total & 4(4) & 4(4) & 4(4) & 2(2) & 4(4) & 6(6) & 6(6) \\

\hline
Number of ``wrong  &\multirow{2}{*}{\large $\frac{4}{a^{2}}$} &\multirow{2}{*}{\large $\frac{4}{a^{2}}$} &\multirow{2}{*}{\large $\frac{4}{a^{2}}$} &\multirow{2}{*}{\large $\frac{4}{\sqrt{2}a^{2}}$} &\multirow{2}{*}{\large $\frac{4}{\sqrt{3}a^{2}}$} &\multirow{2}{*}{\large $\frac{6}{\sqrt{3}a^{2}}$} &\multirow{2}{*}{\large $\frac{6}{\sqrt{3}a^{2}}$} \\
bonds'' per unit area \\
\bottomrule
\end{tabular}
\end{threeparttable}
\end{table*}

\begin{figure}[!htbp]
\centering
\includegraphics*[width=8cm,keepaspectratio]{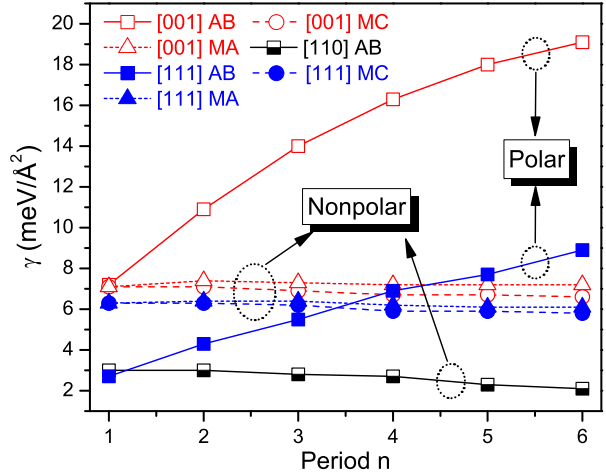}
\caption{\label{fig:f3}(Color online) Interfacial energies of [001] (red empty symbols), [110] (black half-filled symbols), and [111] (blue solid symbols) orientations of (GaSb)$_{n}$(ZnTe)$_{n}$ superlattices as a function of layer period $n$ in forming polar or nonpolar interfaces.}
\end{figure}

Our calculated interfacial energies for different orientations and layer thicknesses for the (GaSb)$_{n}$(ZnTe)$_{n}$ superlattices are shown in Fig. 2. For the nonpolar interface structures (e.g., 50\%-50\% mixed interfaces in the [001] orientation, abrupt interfaces in the [110] orientation, and 25\%-75\% mixed interfaces in the [111] orientation), because the excess electrons and holes can locally and fully compensate each other in a single interface [Figs. 1(b), (c) and (f), (g)], from a Coulomb interaction point of view, they should have the lowest interfacial energy compared with other interfacial structures. However, when the layer thickness $n$ is very small, the charge transfer between the adjacent interfaces and subsequent Coulomb interaction between the interfaces can also stabilize the superlattice. This implies that for the ultrathin III-V/II-VI superlattices, the difference in Coulomb energy derived from the excess carriers between the nonpolar and polar interfaces could be similar. In this case, the stability of the interfacial structures should be mainly determined by the interfacial bond energy, which mainly depends on the number of wrong bonds per unit area near the interfaces in these lattice-matched systems. Table I shows that abrupt [111] interfaces have the lowest number of wrong bonds per unit area, this explains why [111] superlattices have the lowest interface energies for ultrathin superlattices (Fig. 2). As the layer thickness $n$ increases, the Coulomb energy of the polar interfaces will be less attractive due to the increased interface distance, so the interfacial energy increases significantly, as shown for the abrupt interfaces of [001] and [111] in Fig. 2. For nonpolar interfaces, due to the excess charge are fully and locally compensated in each single interface, the Coulomb energy almost does not change when the layer thickness $n$ increases. Because the interfacial bond energy is almost independent of $n$, the interfacial energies of the nonpolar interfaces are also almost independent of $n$ (Fig. 2). For the [111] orientation, as discussed above, for the ultrathin superlattices the polar interfaces are more stable than the nonpolar interfaces. As $n$ increases, the polar interface becomes more and more unstable. When $n\geq4$, the interfacial energy of the polar surface is higher than that of the nonpolar interfaces. This implies that for the long-period superlattices, the Coulomb energy gain is more dominant. For the [110] orientation, the nonpolar surface is always more stable since the abrupt interfaces are already nonpolar. For the [001] orientation, because the formation of mixed nonpolar interfaces doesn't change the number of wrong bonds per unit compared with polar interfaces (Table I), the dominant factor for the interface stability is the Coulomb energy. Because the mixed nonpolar interfaces always have the most negative Coulomb energy, the [001] orientation nonpolar interfaces are always energetically more favorable than the polar interfaces.

\begin{figure}[!ht]
\centering
\includegraphics*[width=11cm,keepaspectratio]{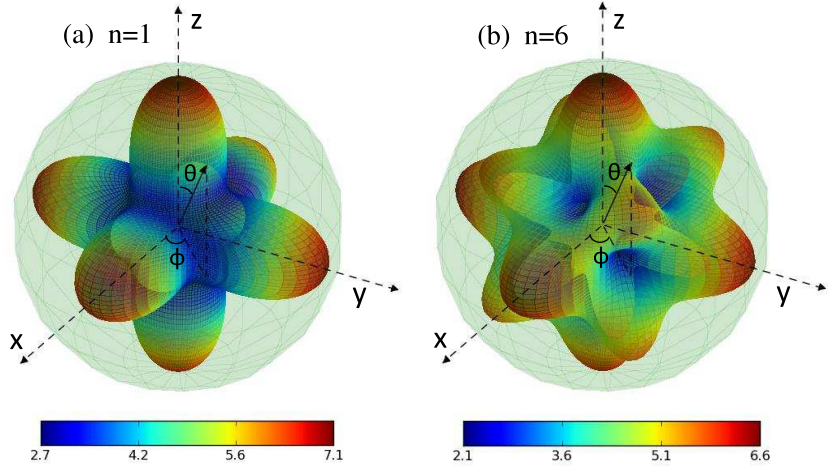}
\caption{\label{fig:f4}(Color online) Interfacial energy (in meV/\AA$^{2}$) of (GaSb)$_{n}$(ZnTe)$_{n}$ superlattices as a function of orientation for periods (a) $n$=1 and (b) $n$=6.}
\end{figure}

Our calculated results show that in the short-period limit ($n$=1), the [111] orientation has the lowest interfacial energy of 2.7 meV/\AA$^{2}$ between these three principal orientations. This is because in the ultrathin limit the stability of the interface is controlled by the number of wrong bonds. Evidently, the minimum number of wrong bonds per unit area for the [001], [110], and [111] orientations is $4/a^{2}$, $4/\sqrt{2}a^{2}$ and $4/\sqrt{3}a^{2} $ (Table I), respectively. Therefore, the interfacial energy has the order of $\gamma_{001}> \gamma_{110}> \gamma_{111}$. As $n$ increases, the interfacial energies of the ground states of the [001] and [110] orientations almost do not change because they both are nonpolar, whereas the interface of the [111] superlattice switches from polar to nonpolar, and the interfacial energy of the ground state of [111] orientation will increase at the beginning, and then change to a constant when the ground state becomes nonpolar.  Thus, for the long-period superlattices, the [001], [110], and [111] orientations all are nonpolar and the interfacial energy has the order of $\gamma_{001}> \gamma_{111}> \gamma_{110}$. That is because the minimum number of wrong bonds per unit interfacial area are $4/a^{2}$, $4/\sqrt{2}a^{2}$, and $6/\sqrt{3}a^{2}$ for the [001], [110] and [111] nonpolar interfaces, respectively.

In order to further confirm the favored growth orientation in the heterovalent superlattices, we need to calculate the interfacial energy along an arbitrary growth orientation as an function of the polar and azimuthal angles ($\theta$, $\phi$). We do this by employing crystal symmetry and expanding the interfacial energy for any orientation ($l,m,n$) or angle ($\theta$, $\phi$) in terms of the cubic lattice harmonics with $l_{max}=6$\cite{von,Yong-Hua_Li}:
\begin{equation}
\begin{aligned}
\gamma(\theta,\phi)&=a+b[\sin(\theta)^{4}\cos(\phi)^{4}+\sin(\theta)^{4}\sin(\phi)^{4}\\
&+\cos(\theta)^{4}-\frac{3}{5}]+c\{\sin(\theta)^{4}\cos(\theta)^{2}\cos(\phi)^{2}\sin(\phi)^{2}\\
&+\frac{1}{22}[\sin(\theta)^{4}\cos(\phi)^{4}+\sin(\theta)^{4}\sin(\phi)^{4}\\
&+\cos(\theta)^{4}-\frac{3}{5}]-\frac{1}{105}\}
\end{aligned}
\end{equation}
where $\theta \in [0,\pi]$ and $\phi \in [0,2\pi)$. The parameters a, b, and c can be fitted using the interfacial energies of the [001] ($\theta=0, \phi=$ arbitrary value), [110] ($\theta=\pi/2, \phi=\pi/4$) and [111] ($\theta=arccos(1/\sqrt{3}), \phi=\pi/4$) directions. Fig. 3 shows the interfacial energy of the (GaSb)$_{n}$(ZnTe)$_{n}$ superlattices as a function of the orientation for (a) $n$=1 and (b) $n$=6. It is found that whether in the short-period limit or in long-period superlattices, the [001] orientation always has the highest interfacial energy. However, for $n$=1 ultrathin superlattices, the minimum interfacial energy is located at the [111] as well as its equivalent orientations. For $n$=6, the [110] and its equivalent orientations have the lowest interfacial energy among all growth orientations. These results suggest that for the short-period superlattices, the [111] and its equivalent orientations is most preferred, and in long-period superlattices, the [110] and its equivalent orientations have the lowest interfacial energy among all growth orientations. On contrary, for isovalent superlattices, the [201] superlattices usually always have the lowest energy \cite{SHWei1989}.

\subsection{IV. CONCLUSIONS}
In conclusion, the trend of stability of chemically mismatched but lattice-matched heterovalent superlattices is significantly different from the chemically matched but lattice-mismatched isovalent superlattices, because for isovalent superlattices the stability is mostly determined by strain, whereas for lattice-matched nonisovalent superlattices the interfacial energy depends not only on the bond energy but also on the Coulomb energy derived from the donor and acceptor wrong bonds. The competition between the bond energy and Coulomb energy determines the structural stability of heterovalent superlattices. In the short-period superlattices, the interfacial bond energy is dominant. That is why the abrupt [111] interface has the lowest energy even though it is polar. However, for the long-period superlattices, the Coulomb energy becomes more dominant, so all the interfaces tend to be nonpolar and the [110] interface has the lowest energy. This chemical trend is applicable to many lattice-matched III-V/II-VI semiconductor superlattices, such as, GaAs/ZnSe, GaP/ZnS, GaN/ZnO, and GaSb/ZnTe, and this understanding provides guidelines for future application of these unique material systems.

\subsection{ACKNOWLEDGMENTS}
The work at Institute of Semiconductors, Chinese Academy of Sciences was supported by the National Basic Research Program of China (973 Program) Grant No. G2009CB929300, and the National Natural Science Foundation of China under Grants No. 61121491, and No. 11104264. The work at NREL was supported by the U.S. Department of Energy under Contract No. DE-AC36-08GO28308.

%


\end{document}